\documentclass[twoside]{article}
\usepackage{fleqn,espcrc2}
\usepackage{epsfig}
\usepackage{graphicx}
\newcommand{\AmS}{{\protect\the\textfont2
  A\kern-.1667em\lower.5ex\hbox{M}\kern-.125emS}}

\title{Testing large mixing
 MSW solutions of the solar neutrino
 problem through Earth regeneration effects}

\author{A. Palazzo \address{Dipartimento di Fisica and Sezione INFN di Bari,
Via Amendola 173, I-70126 Bari, Italy  }}%

\begin{document}

\begin{abstract}
Large mixing MSW solutions to the solar neutrino problem appear
to be currently favored by the data. We discuss the possibility of
discriminating them by means of present and future experiments.
In particular, we show that the study of energy and time dependence
of the Earth regeneration effect can be useful in this respect.
\vspace{1pc}
\end{abstract}

\maketitle

\section{Introduction}

The available neutrino data from the Homestake \cite{Cl}, GALLEX-GNO \cite{GA},
SAGE \cite{SA}, Kamiokande \cite{Kam} and Super-Kamiokande \cite{SK} experiments indicate a
significant $\nu_e$ flux deficit with respect to the predictions of the 
Standard Solar Model (SSM) \cite{BP98}. The $\nu$ oscillation mechanism
represents a possible explanation of this deficit. In particular,
recent analyses \cite{FL_HL} show a preference of the data for the
so-called MSW \cite{MSW} (Mikheyev-Smirnov-Wolfenstein)
solutions involving large mixing, usually indicated
as ``large mixing angle'' (LMA) and ``low $\delta m^2$'' (LOW) solutions.
We point out how present and future experiments can help to
discriminate such two solutions through Earth regeneration effects.

%\section{Discriminating between LMA and LOW solutions by Super-kamiokande
% and SNO}
\section{Energy dependence of the Earth regeneration effect in SK and SNO}

The slightly positive indication ($\simeq 1.3\sigma$) for an excess of
nighttime to daytime events in Super-Kamiokande \cite{SK}, if confirmed with
higher statistical significance, would indicate the occurence of the Earth
regeneration effect for $^8B$ neutrinos. Such indication, by itself, might not
be sufficient to discriminate the LMA from the LOW solution, since a slight
excess is predicted in both cases. On the other hand, the Earth regeneration
effect depends strongly on the neutrino energy and,
in principle, one could take advantage of this
feature to discriminate the two solutions. In
particular, starting from the simple observation that the Earth regeneration
effect is stronger at low energy for the LOW solution, and at high energy for
the LMA solution, we point out that it may be useful to study the night-day
asymmetry in two separate energy ranges in both the Super-Kamiokande and the 
SNO experiments \cite{FL_HL}.
For definiteness, we consider the two following representative ranges for the
total (measured) energy of recoiling electrons in SK and SNO,

%-----------------------------------------------------------
\begin{equation}
{\rm Low\,\, range\,(L)} = [5.0, 7.5]\,{\rm MeV}
\label{eq:Lowrange}
\,,
\end{equation}
%-----------------------------------------------------------
\begin{equation}
{\rm High\,\, range\,(H)} = [7.5, 20]\,{\rm MeV}
\label{eq:Highrange}
\,,
\end{equation}
%-----------------------------------------------------------
and calculate the night-day rate asymmetry in such ranges,
%-----------------------------------------------------------
\begin{equation}
A_{H,L} = \left(\frac{N-D}{N+D}\right)_{H,L}
\,.
\end{equation}
%-----------------------------------------------------------
Since one expects $A_H > A_L$ for the LMA solution and $A_H < A_L$ for the LOW
solution, it is useful to introduce the difference
%-----------------------------------------------------------
\begin{equation}
\Delta = A_H - A_L
\,,
\end{equation}
%-----------------------------------------------------------
which should change sign when passing from the LMA region ($\Delta >0$) to the
LOW region ($\Delta<0$).

\begin{figure}[t!]
\vspace{30pt}
\mbox{
\psfig{bbllx=1.5truecm,bblly=5.0truecm,bburx=19.0truecm,bbury=23.1truecm,
height=7.0truecm,figure=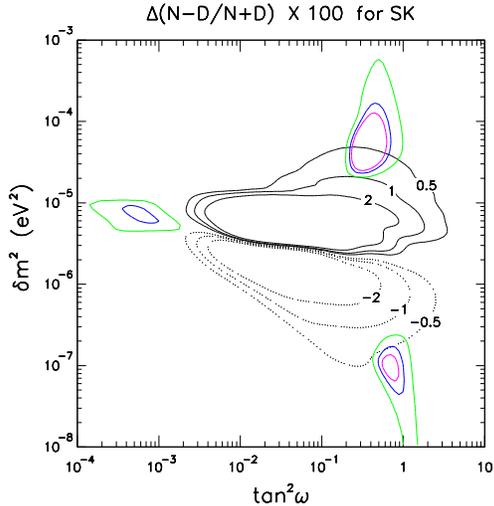}}
%\fantom{.}
\vspace{-55pt}
\caption{\label{fig:SK}
Curves of iso-$\Delta\times100$ for Super-Kamiokande. The MSW
 solutions at $90\%$, $95\%$, $99\%$ are also drawn.}
\vspace*{-15pt}
\end{figure}

\begin{figure}[hbt]
\vspace{30pt}
\mbox{
\psfig{bbllx=1.5truecm,bblly=5.0truecm,bburx=19.0truecm,bbury=23.1truecm,
height=6.8truecm,figure=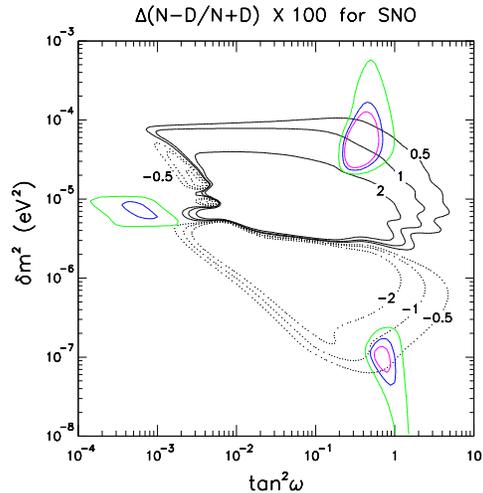}}
\vspace{-50pt}
\caption{\label{fig:SNO}Curves of iso-$\Delta\times100$ for SNO.}
\end{figure}

Figures  \ref{fig:SK} and  \ref{fig:SNO} show the result of our
calculations of $\Delta$ (eccentricity effects removed) in SK and SNO,
respectively, in the form of isolines at $\Delta \times 100 = \pm0.5$,
$\pm1$ and $\pm2$, superimposed to the current MSW solutions at $90\%$, $95\%$
and $99\%$ C.L., obtained using the latest experimental data.
Such figures confirm that $\Delta$ sign effectively
changes when passing from LMA to LOW solution, thus
representing a useful test to discriminate the two large mixing
angle solutions.

As one can notice in the figures  \ref{fig:SK} and  \ref{fig:SNO}, it turns out
that, in the most favorable circumstances, the value of $\Delta \times
100$ can reach $-0.5$ (SK) and $-1.0$ (SNO) in the case of the LOW solution
and 1.0 (SK) and 2.0 (SNO) for the LMA. This small effect could be observable
in the two experiments, provided that they can reach a sensitivity to
$\Delta$ at a (sub)percent level. However, the simple preference of the
SK and SNO experiments for a definite $\Delta$ sign would already constitute a
useful information, corroborating other tests envisaged to
solve the LOW-LMA ambiguity \cite{SMI1,SMI2,SMI3}.

%-----------------------------------------------------------
%-----------------------------------------------------------
Our choice of the low and high energy ranges in Eqs.
(\ref{eq:Lowrange}, \ref{eq:Highrange})  is only
representative, and it should be tuned to reach the best compromise
between sensitivity to the energy dependence of Earth regeneration effect
and statistical significance. For instance, one could also
enhance the sensitivity to the energy dependence by calculating
the day-night asymmetry in a larger number of bins 
%-----------------------------------------------------------
%-----------------------------------------------------------
 (D-N asymmetry spectrum); however, the statistical significance would
decrease in each bin. The optimal compromise requires dedicated experimental
studies.

\section{Testing LOW solution through time variations in BOREXINO and GNO}
The Earth regeneration effect occurs at relatively low values of
$\delta m^2$ for low energy ($^7$Be and pp) neutrinos, that can be
detected by the two experiments BOREXINO (based on $\nu_e$ scattering) 
and GNO (based on $\nu_e$-{\rm Ga} absorption). The first is a real-time
experiment, able to detect a possible regeneration phenomenon as a day-night
asymmetry of the event rate. The second is a radiochemical experiment with a
signal extraction timescale of $\sim 1$ month. It can, however, detect
Earth effect in the form of a seasonal modulation of the
signal \cite{FL_GNO}. Indeed, during winter, the nighttime is
longer compared to summer and, furthermore, the trajectory of
neutrinos probes inner layers of the Earth (in particular, the Earth core
is crossed only during winter \cite{FL_GNO}).
%-----------------------------------------------------------
\begin{figure}[t!]
\vspace{30pt}
\mbox{
\psfig{bbllx=1.5truecm,bblly=5.0truecm,bburx=19.0truecm,bbury=23.1truecm,
height=6.8truecm,figure=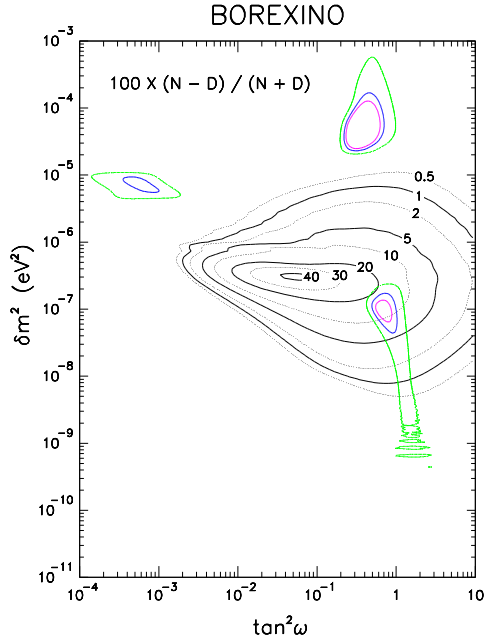}}
\vspace{-2pt}
\caption{\label{fig:BOR}Curves of iso-$A_{DN}\times100$ for BOREXINO.
The current MSW+quasivacuum solutions at $90\%$, $95\%$, $99\%$
are also drawn.}
\end{figure}
%%-----------------------------------------------------------
\begin{figure}[hbt]
\vspace{10pt}
\mbox{
\psfig{bbllx=1.5truecm,bblly=5.0truecm,bburx=19.0truecm,bbury=23.1truecm,
height=6.8truecm,figure=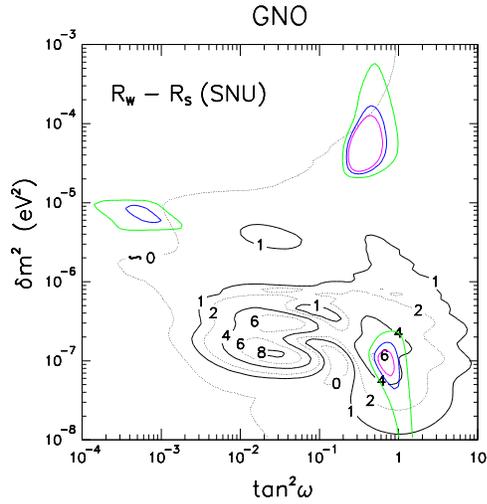}}
\vspace{-30pt}
\caption{\label{fig:GNO}Curves of iso-$R_{WS}$ (in SNU) for GNO.}
\end{figure}
%-----------------------------------------------------------

In Fig. \ref{fig:BOR} we show the results of our calculation of
the day-night asymmetry ($A_{DN}\times 100$) of the event rate for BOREXINO.
In the same plot we show the current solutions of the solar neutrino
problem at $90\%$, $95\%$ and
$99\%$ C.L. (for 2 d.o.f.) respectively. This figure shows that the
day-night asymmetry expected in BOREXINO in the region of the LOW
solution is expected to be in the range $1\%-20\%$.

Figure \ref{fig:GNO} shows the iso-lines of Winter-Summer rate
difference $R_{WS}$ (in SNU) expected in GNO due to the Earth
regeneration effect (eccentricity effect removed). Winter and
Summer periods are defined as
\begin{equation}
{\rm Winter} ~\,\simeq [23\ {\rm september},  21\ {\rm  march}]
\,,
\end{equation}
\begin{equation}
{\rm Summer} \simeq [22\ {\rm march}, 22\ {\rm september}]
\,.
\end{equation}
Notice that the
sensitivity in the LOW region is almost maximal since GNO is
predominantly sensitive to low-energy pp $\nu$'s, as compared with BOREXINO,
which is more sensitive to the $^7$Be line $\nu$'s. This circumstance makes
these two experiments intrinsically different in testing the LOW
solution. The two tests are also complementary in tracking the origin of the
time variations: a possible seasonal signal in GNO, if originated by vacuum
(instead of MSW) oscillations, should not produce a N-D asymmetry in BOREXINO.

%\section{conclusion}
%We have showed how is possible to discriminate the large mixing solutions
%to the solar neutrino problem by the help of the Earth regeneration effect.
%Infact SuperKamiokande and SNO could be able to evindence an eventual

\end{document}